\documentclass[a4paper]{jpconf}
\usepackage{graphicx}
\begin{document}
\title{Recent ALICE results on quarkonium production in nuclear collisions}

\author{Biswarup Paul~\footnote{on behalf of the ALICE Collaboration}}

\address{University and INFN Cagliari,\\
  Cittadella Universitaria di Monserrato, 09042, Monserrato (CA), Italy}

\ead{biswarup.paul@cern.ch}

\begin{abstract}
Quarkonium production has long been regarded as a potential signature of deconfinement in nucleus-nucleus collisions. Recently, the production of J/$\psi$ via regeneration within the quark-gluon plasma (QGP) or at the phase boundary has been identified as an important ingredient for the interpretation of quarkonium production results from lead-lead collisions at the Large Hadron Collider (LHC). Quarkonium polarization could also be used to investigate the properties of the hot and dense medium created at the LHC energies. In this contribution, the latest ALICE results on quarkonium will be presented and discussed. These include, among others, the nuclear modifications of (prompt and non-prompt) J/$\psi$ and $\psi$(2S) production, and the J/$\psi$ polarisation, all measured with lead-lead collisions at the LHC. The results will be compared with available theoretical model calculations.
\end{abstract}

\vspace{-0.6cm}
\section{Introduction}
The characterization of the QGP is the main goal of ultra-relativistic heavy-ion collision studies. Charmonium is one of the most prominent probes used to investigate and quantify the properties of the QGP. The study of the excited $\psi$(2S) state is of particular interest. Because of the larger size and weaker binding energy of the $\psi$(2S) state, the effects of nuclear medium on its production might be significantly different from those of the J/$\psi$. The regeneration mechanism, within the QGP or at the phase boundary, is an important ingredient for describing the J/$\psi$ production at the LHC energies. $\psi$(2S) production relative to J/$\psi$ represents one possible discriminator between the two different regeneration scenarios. 
The difference of the J/$\psi$ polarisation between Pb--Pb~\cite{pol_old} and pp collisions~\cite{pol_LHCb} at LHC could be related to the modification of the J/$\psi$ feed down fractions, due to the suppression of the excited charmonium states in the QGP, but also to the contribution of the regenerated J/$\psi$ in the low transverse momentum ($p_{\rm T}$) region. Moreover, it has been hypothesized that quarkonium states could be polarized by the strong magnetic field, generated in the early phase of the evolution of the system, and by the large angular momentum of the medium in non-central heavy-ion collisions. This kind of information can be assessed by defining an ad hoc reference frame where the quantization axis is orthogonal to the event plane of the collision.

\vspace{-0.2cm}
\section{ALICE detector and data samples}
The ALICE experiment has studied inclusive J/$\psi$ and $\psi$(2S) production in Pb--Pb collisions at $\sqrt{s}_{\rm NN} = \mbox{5.02 TeV}$ through its dimuon decay channel. Muons are identified and tracked in the Muon Spectrometer, which covers the pseudorapidity range $-4<\eta<-2.5$~\cite{alice}. The pixel layers of the Inner Tracking System (ITS) allow the vertex determination, while forward VZERO scintillators are used for triggering purposes. The VZERO is also used to determine the centrality of the collisions. At midrapidity ($|y| < 0.9$), J/$\psi$ reconstruction is done via dielectron decay channel down to $p_{\rm T}$ = 0 GeV/$c$. The track reconstruction is performed by the central barrel detectors, especially Time Projection Chamber (TPC) and ITS. Particle identification is based on the measurement of specific energy loss in the active volume of TPC, while primary and secondary vertex reconstruction is done by the innermost layers of the ITS. Prompt and non-prompt J/$\psi$ separation is possible down to $p_{\rm T}$ = 1.5 GeV/$c$ at midrapidity in Pb--Pb collisions. 

\vspace{-0.2cm}
\section{Results}
\begin{figure}[ht]
\centering
\includegraphics[scale=0.39]{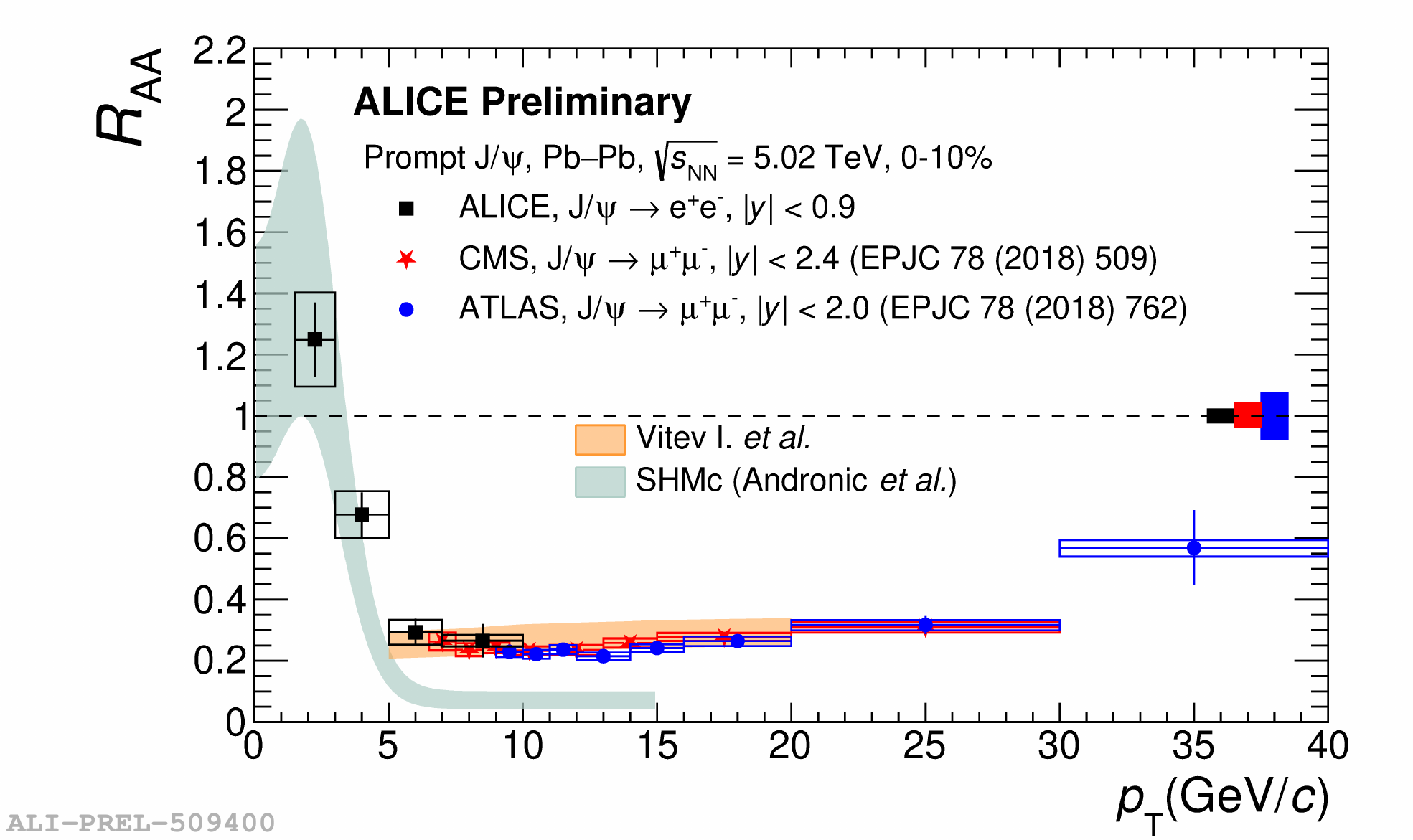}
\includegraphics[scale=0.39]{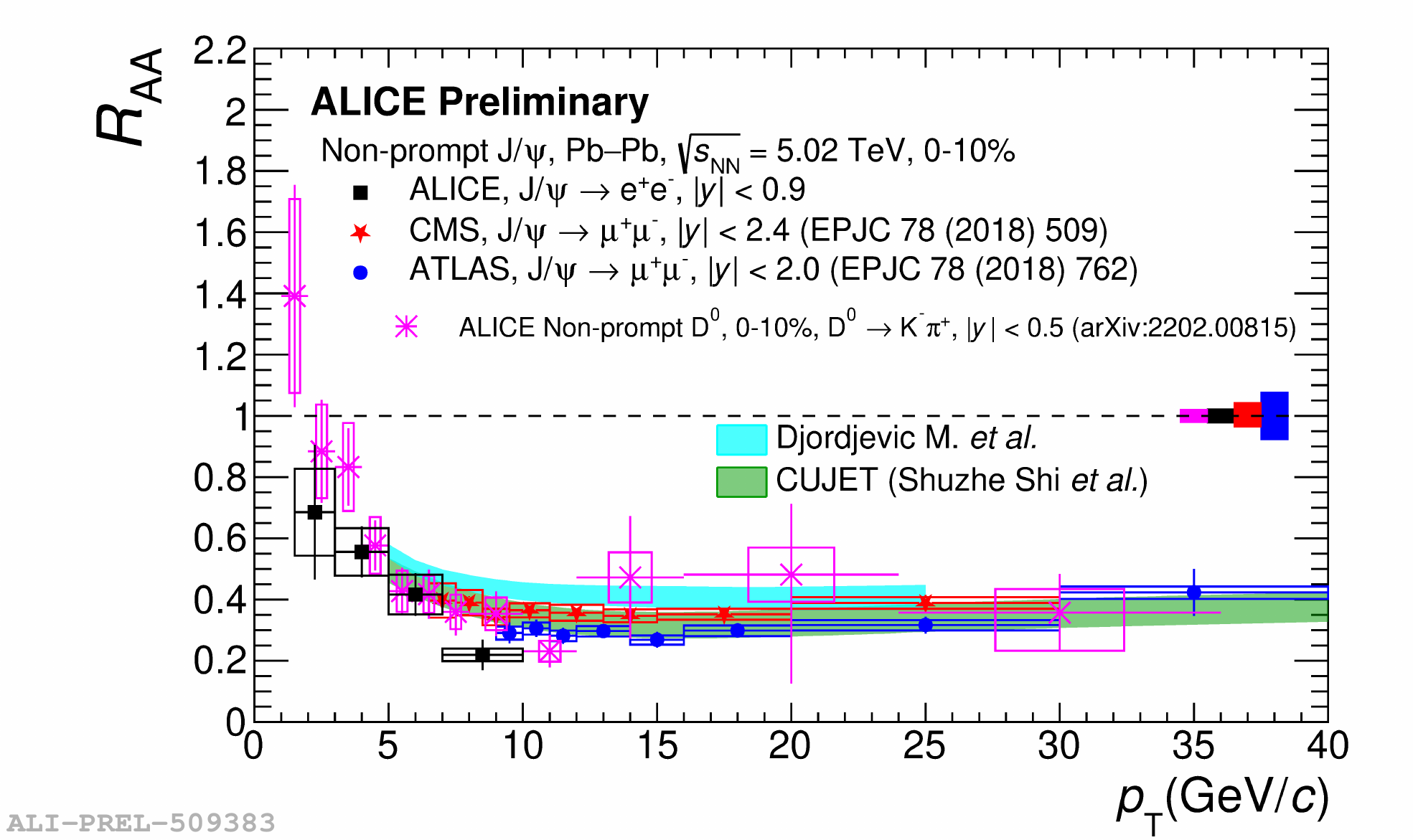}
\vspace{-0.2cm}
\caption{\label{Raaprompt}$R_{\rm AA}$ of prompt (left) and non-prompt (right) J/$\psi$ as a function of $p_{\rm T}$ in 0--10\% central Pb--Pb collisions at $\sqrt{s}_{\rm NN} = \mbox{5.02 TeV}$ TeV at midrapidity compared to CMS and ATLAS results and to the models.}
\end{figure}

The prompt and non-prompt J/$\psi$ $R_{\rm AA}$ at midrapidity in 0--10\% centrality is shown in Fig.~\ref{Raaprompt}. The results are in consistent with CMS and ATLAS measurements at high $p_{\rm T}$ in 0--10\% central collisions. The statistical hadronization model (SHMc)~\cite{SHMc,SHMc2} for charm quark including J/$\psi$ regeneration at the phase boundary, reproduces the prompt J/$\psi$ $R_{\rm AA}$ at low $p_{\rm T}$ ($p_{\rm T} < 5$ GeV/$c$). Models including J/$\psi$ dissociation and charm quark energy loss in medium~\cite{Ivan,Saron} are consistent with prompt J/$\psi$ $R_{\rm AA}$ at $p_{\rm T}$ larger than 5 GeV/$c$. The non-prompt J/$\psi$ $R_{\rm AA}$ is consistent with the non-prompt D$^{0}$ $R_{\rm AA}$ and models including radiative and collisional energy losses for b quarks~\cite{Shi,Shi2,Zigic} describe the non-prompt J/$\psi$ $R_{\rm AA}$ within uncertainties.

\vspace{-0.2cm}

\begin{figure}[ht]
\centering
\includegraphics[scale=0.38]{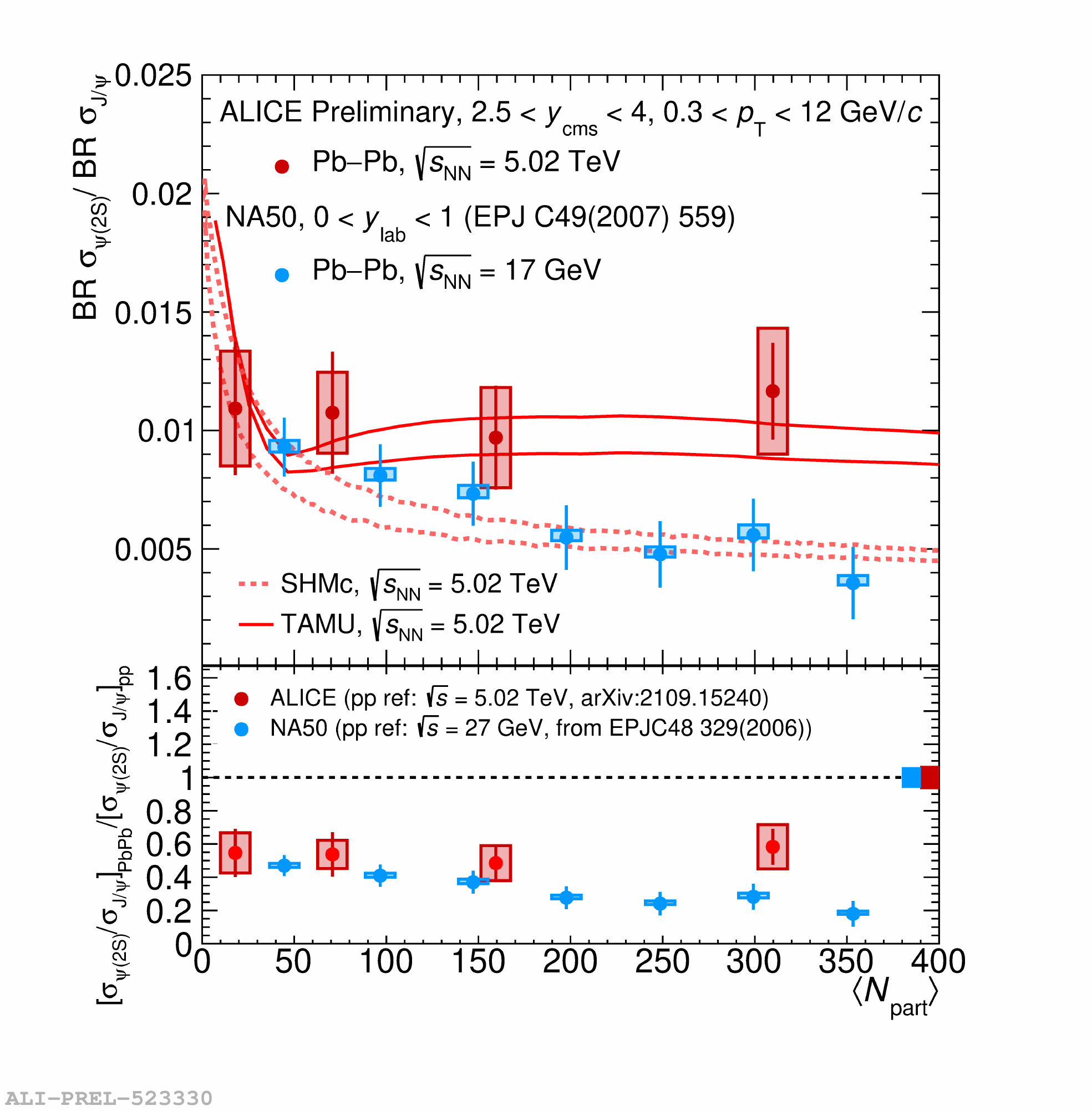}
\includegraphics[scale=0.38]{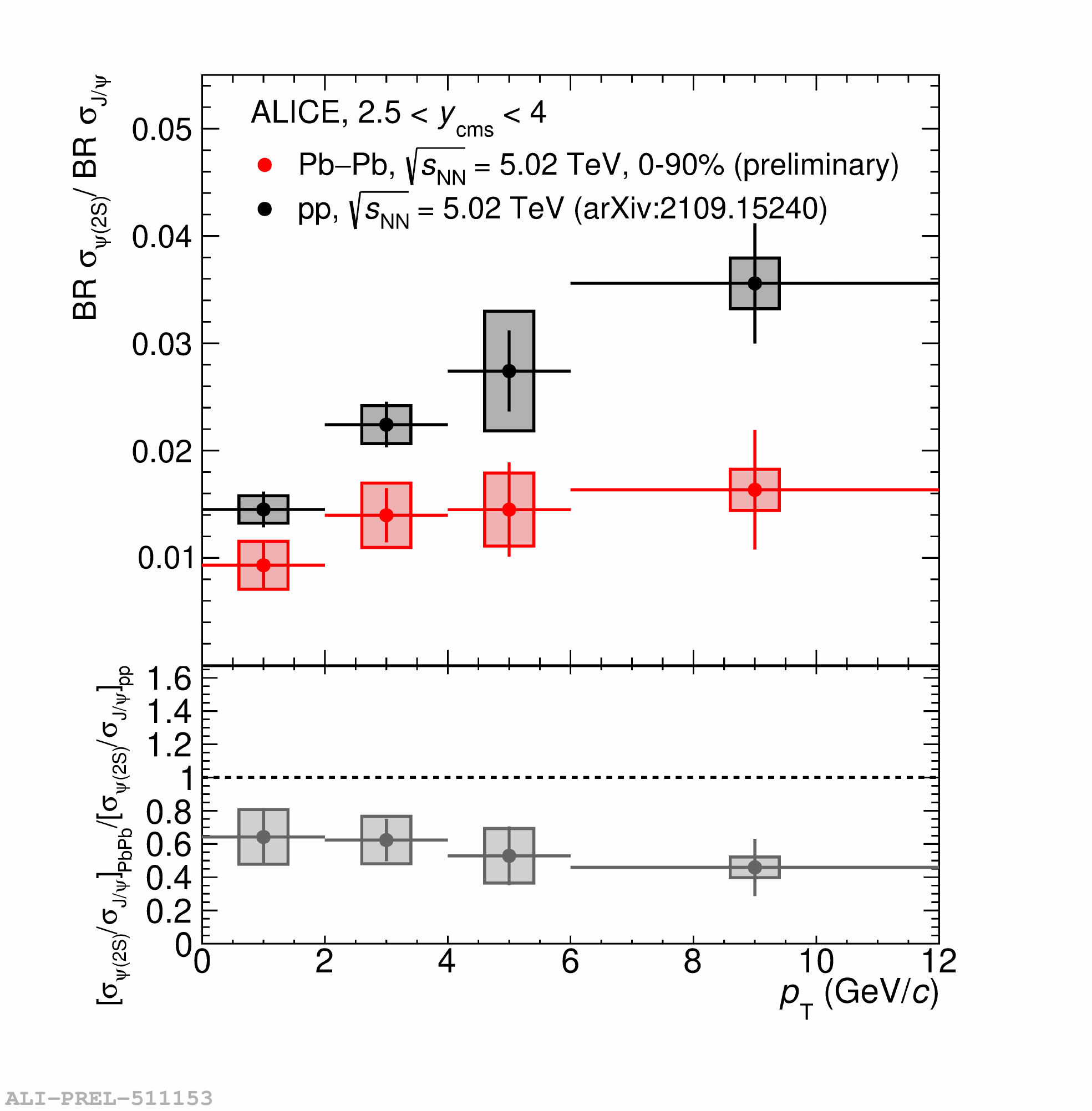}
\vspace{-0.2cm}
\caption{\label{Psi2SJPsiratios}$\psi$(2S)-to-J/$\psi$ cross section ratio measured by the ALICE collaboration in Pb--Pb collisions at $\sqrt{s_{\rm NN}}$ = 5.02 TeV as a function of $\langle N_{\rm part}\rangle$ (left) and $p_{\rm T}$ (right). In the left panel, NA50 measurements at SPS carried out at $\sqrt{s_{\rm NN}}$ = 17 GeV are also shown. The results, are compared with theoretical predictions from TAMU~\cite{TAMU} and SHMc~\cite{SHMc,SHMc2}. Bottom panels show the $\psi$(2S)-to-J/$\psi$ ratio normalized to the corresponding pp value (double ratio).}
\end{figure}

The $\psi$(2S)-to-J/$\psi$ cross section ratio measured by the ALICE collaboration in Pb--Pb collisions at $\sqrt{s}_{\rm NN} = \mbox{5.02 TeV}$ at forward rapidity as function of centrality, expressed in terms of average number of participant nucleons $\langle N_{\rm part}\rangle$, is shown in left panel of Fig.~\ref{Psi2SJPsiratios}. The $\psi$(2S)-to-J/$\psi$ double ratio is shown in the bottom panel of Fig.~\ref{Psi2SJPsiratios}, indicating a suppression effect by 50\% in Pb--Pb with respect to pp collisions. Flat centrality dependence is observed within uncertainties. The centrality dependence of both the ratios are compared with NA50 results in Pb--Pb collisions at $\sqrt{s}_{\rm NN}$ = 17 GeV in 0 $< y_{\rm Lab} < 1$ and NA50 exhibit a stronger centrality dependence, reaching smaller values in central collisions. TAMU model~\cite{TAMU} reproduces the centrality dependence of $\psi$(2S)-to-J/$\psi$ ratio, while SHMc~\cite{SHMc,SHMc2} tends to underestimate the result in central Pb--Pb collisions.

In the right panel of Fig.~\ref{Psi2SJPsiratios}, the $p_{\rm T}$ dependence of $\psi$(2S)-to-J/$\psi$ ratio in Pb--Pb collisions is compared with the corresponding ratio in pp collisions. The $\psi$(2S)-to-J/$\psi$ ratio in Pb--Pb collisions is systematically larger compared to the one measured in pp. The corresponding double ratio shown in the bottom panel, indicates a significant relative suppression in Pb--Pb with respect to pp, with no strong $p_{\rm T}$ dependence and reaching a value of $\sim$ 0.5 at high $p_{\rm T}$.

\begin{figure}[ht]
\centering
\includegraphics[scale=0.38]{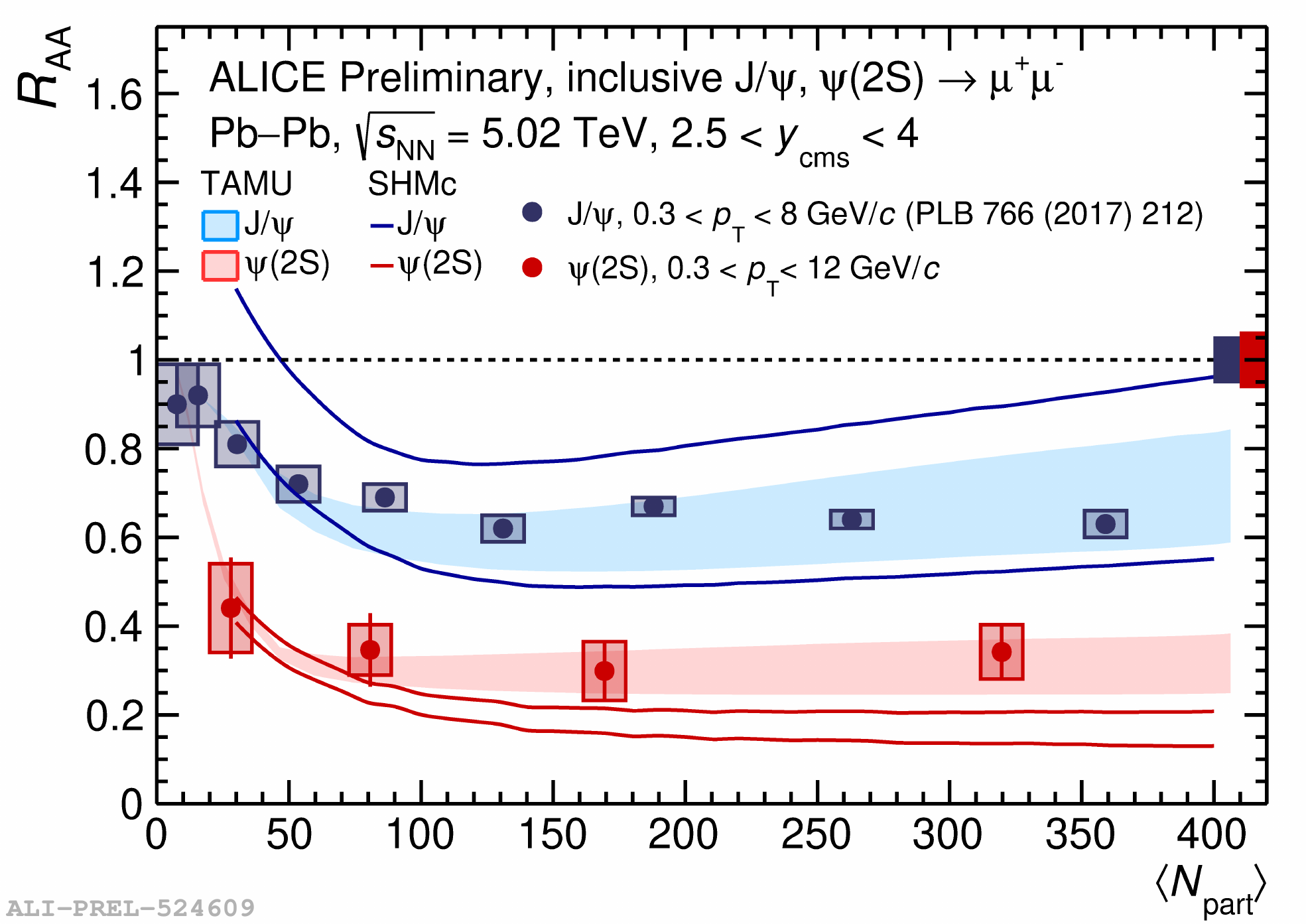}
\includegraphics[scale=0.38]{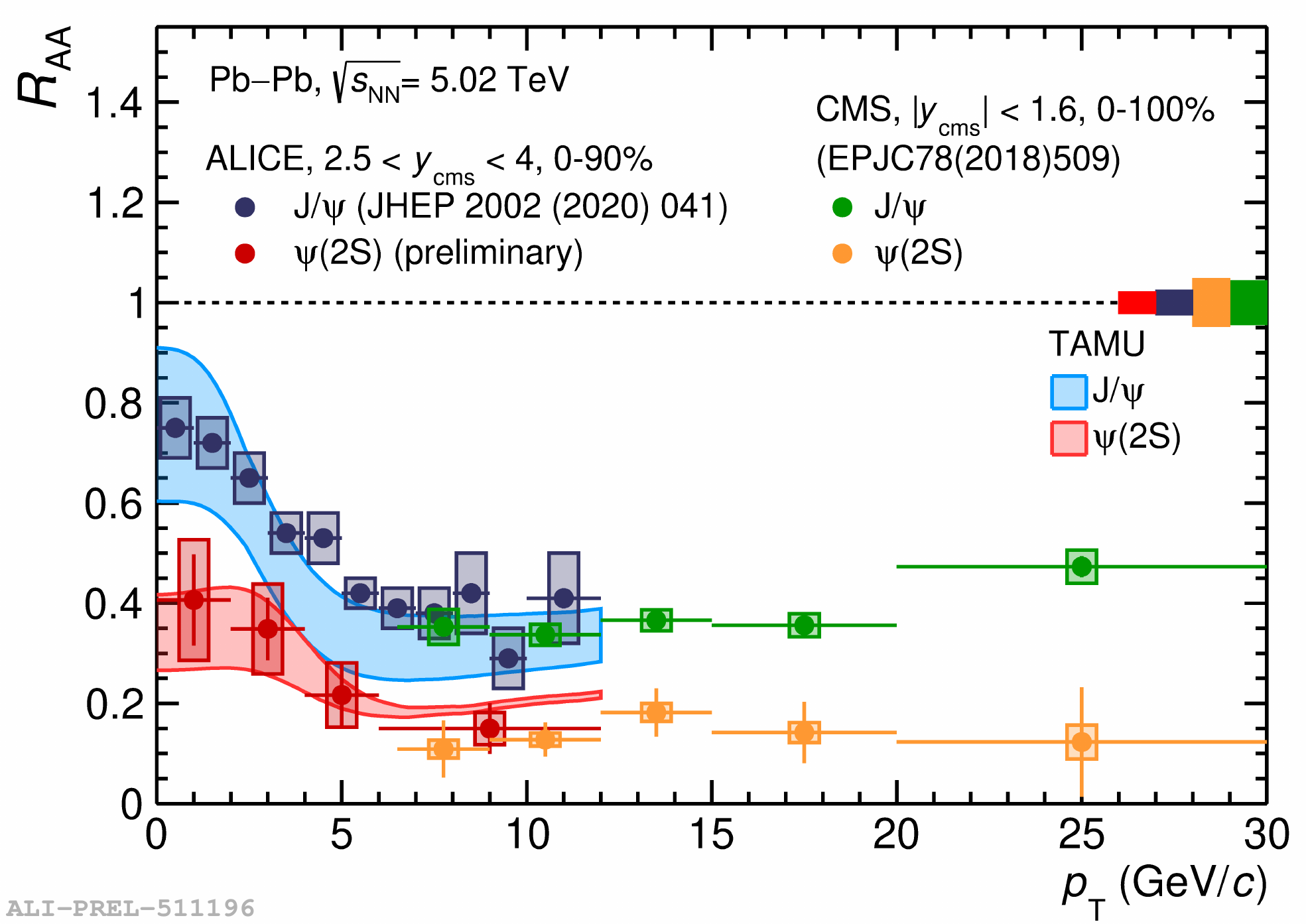}
\vspace{-0.2cm}
\caption{\label{Psi2SJPsiRAA}The $R_{\rm AA}$ of $\psi$(2S) and J/$\psi$ as a function of $\langle N_{\rm part}\rangle$ (left) and $p_{\rm T}$ (right). In the right panel, the ALICE data are compared with CMS results for $|y|$ $<$ 1.6, 6.5 $<$ $p_{\rm T}$ $<$ 30 GeV/$c$ and centrality 0--100 \%. The results are also compared with theoretical predictions from TAMU~\cite{TAMU} (left and right plots) and SHMc~\cite{SHMc,SHMc2} (left plot).}
\end{figure}

Figure~\ref{Psi2SJPsiRAA} shows the nuclear modification factor $R_{\rm AA}$ of J/$\psi$ and $\psi$(2S) measured by the ALICE collaboration as a function of $\langle N_{\rm part}\rangle$ (left panel) and $p_{\rm T}$ (right panel). The results show that $\psi$(2S) is strongly suppressed than J/$\psi$ both as a function of $p_{\rm T}$ and centrality. Flat centrality dependence of $\psi$(2S) $R_{\rm AA}$ is consistent with an $R_{\rm AA}$ value of about 0.4. TAMU model~\cite{TAMU} reproduces the centrality dependence of $R_{\rm AA}$ for both J/$\psi$ and $\psi$(2S), while SHMc~\cite{SHMc,SHMc2} tends to underestimate the $\psi$(2S) $R_{\rm AA}$ in central Pb--Pb collisions. The $p_{\rm T}$ dependence of $R_{\rm AA}$ shows a stronger suppression at high $p_{\rm T}$ and increasing trend of $R_{\rm AA}$ towards low $p_{\rm T}$ for both charmonium states. This is a hint of charmonium regeneration. The result is in good agreement with CMS results for $|y|$ $<$ 1.6, 6.5 $<$ $p_{\rm T}$ $<$ 30 GeV/$c$ and centrality 0--100 \% at high $p_{\rm T}$. TAMU~\cite{TAMU} model reproduces the $p_{\rm T}$ dependence of $R_{\rm AA}$ for both J/$\psi$ and $\psi$(2S).

\begin{figure}[ht]
\centering
\includegraphics[scale=0.38]{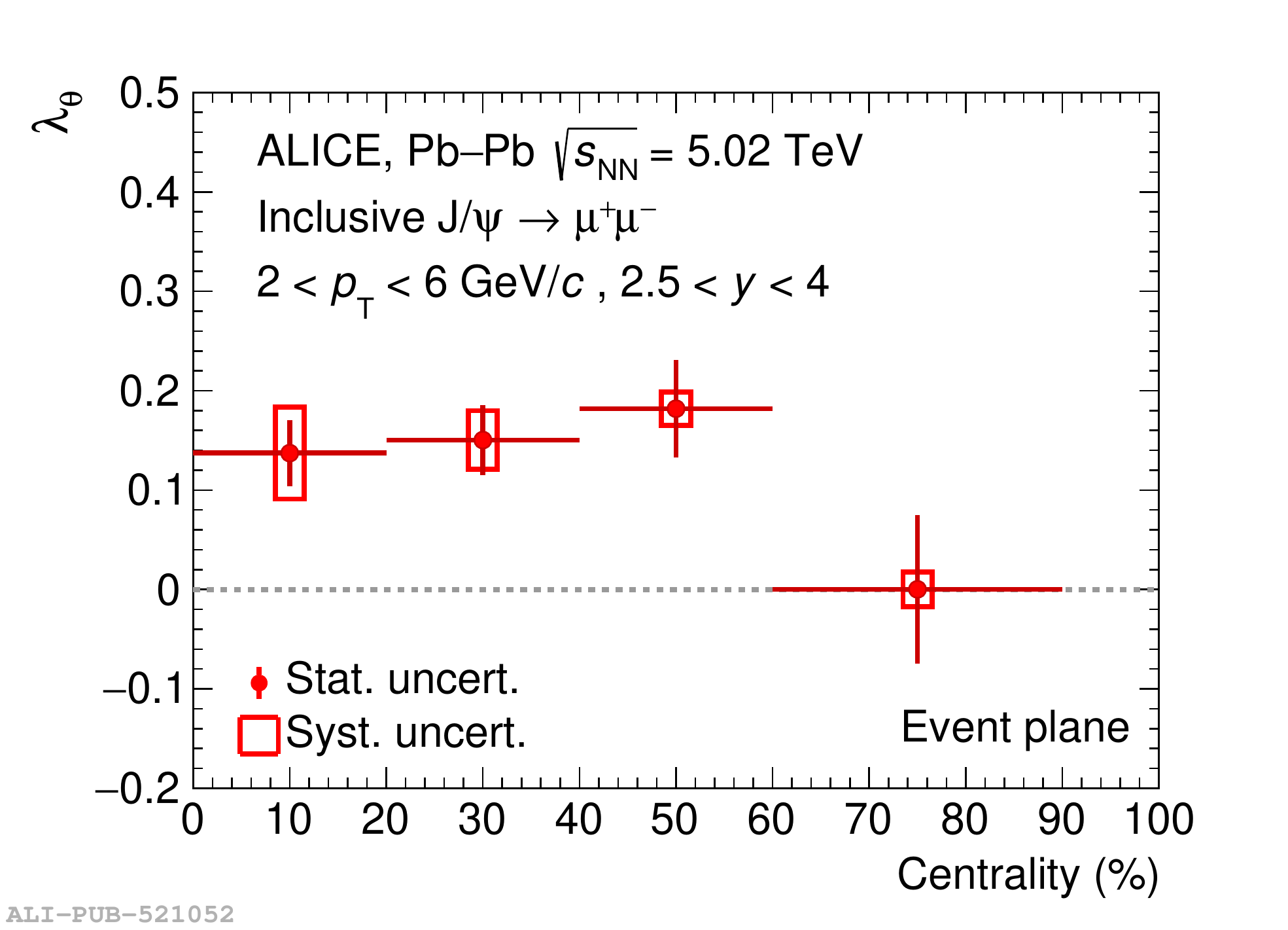}
\includegraphics[scale=0.38]{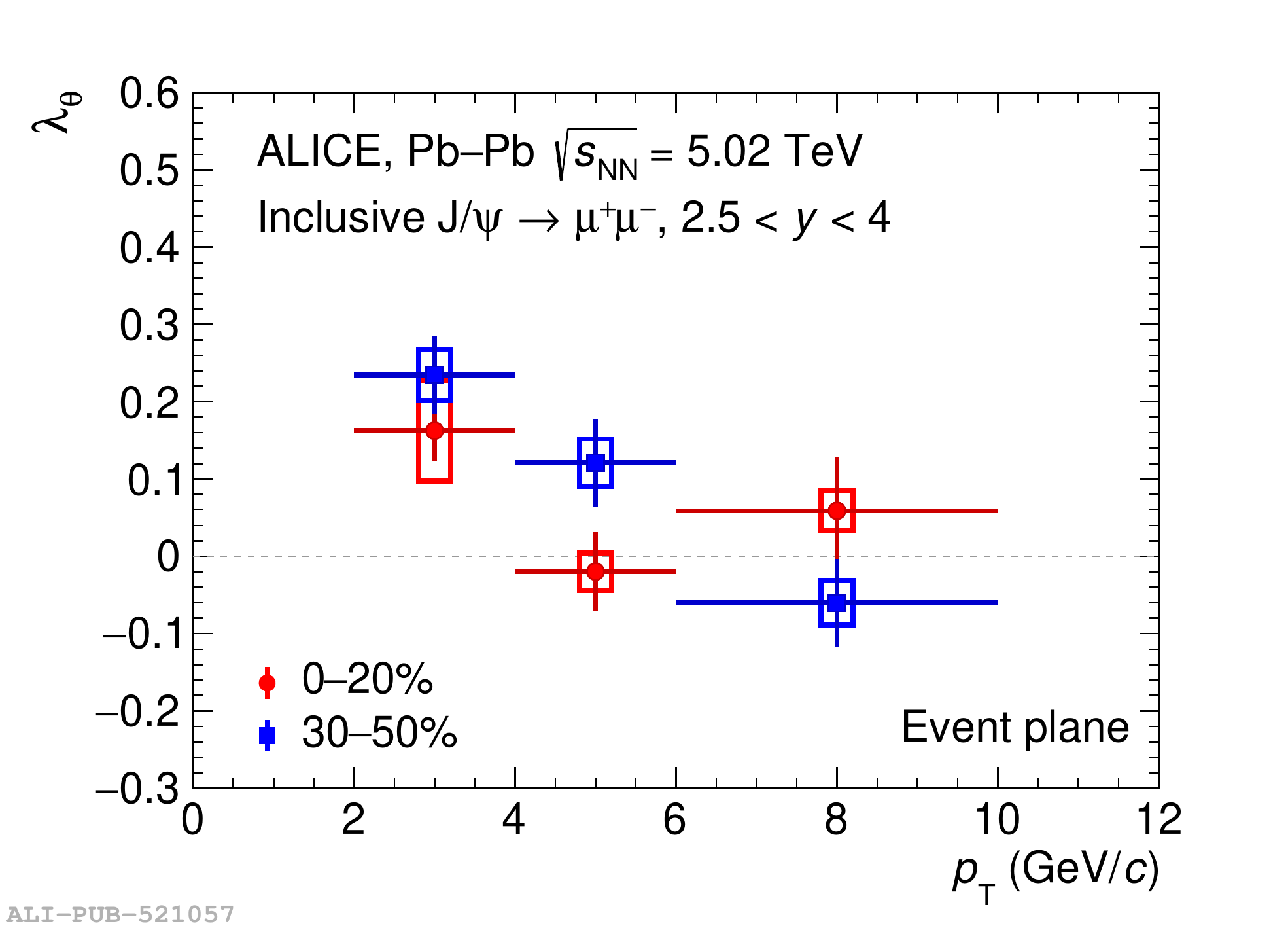}
\vspace{-0.2cm}
\caption{\label{pol_jpsi}Centrality (left panel) and $p_{\rm T}$ dependence (right panel) of $\lambda_{\theta}$ for the J/$\psi$ measured with respect to the axis orthogonal to the event plane in Pb--Pb collisions at $\sqrt{s_{\rm NN}}$ = 5.02 TeV in the forward rapidity region (2.5 $< y <$ 4)~\cite{pol_new}. The vertical bars represent the statistical uncertainties, while the boxes correspond to the systematic uncertainties.}
\end{figure}

J/$\psi$ polarization measurement has been recently extended to a reference frame where the quantization axis corresponds to the direction orthogonal to the event plane (i.e. the plane identified by the impact parameter of the collision and the beam axis)~\cite{pol_new}. This allows one to investigate potential effects related to the magnetic field due to the spectator nucleons~\cite{pol_mag} and the large angular momentum associated to the rotation of the medium produced in the collision~\cite{pol_rot}. The results shown in Fig.~\ref{pol_jpsi} exhibit a maximum deviation of $\sim$ 3.9$\sigma$ with respect to $\lambda_{\theta}$ = 0 in semi-central (30-50\%) collisions for 2 $< p_{\rm T} <$ 4 GeV/$c$. The absence of theoretical predictions prevents from drawing a definitive conclusion.

\vspace{-0.6cm}
\section{Summary}
The nuclear modification factor of prompt and non-prompt J/$\psi$ are measured at midrapidity in Pb--Pb collisions at $\sqrt{s_{\rm NN}}$ = 5.02 TeV. Prompt J/$\psi$ $R_{\rm AA}$ result is described by model including regeneration at low $p_{\rm T}$ and dissociation at high $p_{\rm T}$. Strong suppression for non-prompt J/$\psi$ at high $p_{\rm T}$ is described by models implementing parton energy loss in medium. The first accurate measurement of the $\psi$(2S) production in Pb--Pb collisions at $\sqrt{s_{\rm NN}}$ = 5.02 TeV has been reported by ALICE at forward rapidity. The $\psi$(2S) is more suppressed than the J/$\psi$ as a function of $p_{\rm T}$ and centrality. Transport model (TAMU), which includes recombination of charm quarks in the QGP phase, reproduces the $\psi$(2S) $R_{\rm AA}$ and $\psi$(2S)-to-J/$\psi$ ratio better than SHMc model for central events. Significant J/$\psi$ polarization w.r.t. event plane is observed in Pb--Pb collisions, inputs from theory needed to understand the measurements.

\end{document}